\documentclass[12pt]{iopart}

\usepackage{setstack} % al posto di amsmath
\usepackage{iopams}
\usepackage[dvips]{graphicx}

\newcommand{\mbf}[1]{\ensuremath{\boldsymbol{#1}}}
\newcommand{\RMD}{\ensuremath{\mathrm{d}}}
\DeclareRobustCommand{\Eqref}[1]{Eq.~(\ref{#1})}
\DeclareRobustCommand{\eqref}[1]{(\ref{#1})}
\DeclareRobustCommand{\Figref}[1]{Fig.~(\ref{#1})}
\DeclareRobustCommand{\Secref}[1]{Sec.~\ref{#1}}
\DeclareRobustCommand{\Appref}[1]{\ref{#1}}

\begin{document}

\title{Variations on the magnetic torque acting on a wire}
\author{Claudio Bonati}
\address{Dipartimento di Fisica, Universit\`a di Pisa and INFN, Sezione di Pisa, Largo Pontecorvo 3, 56127 Pisa, Italy.}
\ead{bonati@df.unipi.it}
\date{\today}

\begin{abstract}

The relation $\mbf{M}=\mbf{\mu}\times\mbf{B}$ is presented in all elementary courses on electromagnetism but it is
usually given just for the simple case of a rectangular wire. We will present a completely general but elementary proof
of this relation together with two more advanced proof methods. We will then provide some extensions: non-closed
wires and non-uniform magnetic field. 

\end{abstract}


\maketitle

\section{Introduction}\label{intro_sec}

The torque $\mbf{M}$ acting on a wire induced by a uniform magnetic field $\mbf{B}$ is given by the well known 
formula
\begin{equation}\label{torque}
\mbf{M}=\mbf{\mu}\times\mbf{B}\ ,
\end{equation}
where $\mbf{\mu}$ is the magnetic dipole moment of the wire. In all elementary textbooks on electromagnetism (see e.g. 
Refs.~\cite{griffiths, feynman}) this formula is introduced by studying the example of a rectangular wire, for which 
the magnetic dipole moment can be explicitly written as $AI\mbf{n}$, where $A$ is the area of the wire, 
$I$ is the current passing through it and $\mbf{n}$ is the normal to the plane of the wire, with orientation 
consistent with that of the current.

Also in more advanced textbooks the complete derivation of \Eqref{torque} is usually skipped and just the example of the 
rectangular wires is presented. Two notable exceptions to this rule are the 8th volume of the Landau theoretical physics 
course (Ref.~\cite{landau8}, pag. 128) and the book by J.~D.~Jackson (Ref.~\cite{jackson}, pag. 188-190): the (sketch of 
the) proof by Landau relies on a variation on the theme of the Stokes theorem
\begin{equation}\label{landau_id}
\oint\RMD\mbf{\ell}\times\mbf{X}=
\int\RMD\mbf{\sigma}\times (\mbf{\nabla}\times\mbf{X})+
\int(\RMD\mbf{\sigma}\cdot\mbf{\nabla})\mbf{X} -\int \RMD\mbf{\sigma} (\mbf{\nabla}\cdot\mbf{X})\ ,
\end{equation}
where $\mbf{X}$ is a generic vector field and the integrals in the left and right hand side of the equation are
line and surface integrals respectively. Jackson's proof uses instead the identity
\begin{equation}\label{jackson_id}
\int (f \mbf{Y}\cdot\mbf{\nabla}g+g\mbf{Y}\cdot\mbf{\nabla}f+fg\mbf{\nabla}\cdot\mbf{Y} )\RMD^3 \mbf{r}=0
\end{equation}
where $f$ and $g$ are scalar functions of the position and $\mbf{Y}$ is a vector field of compact support.

We will present a completely elementary proof of \Eqref{torque} together with the ones by Landau and Jackson, reviewed here 
with some more details than in the original references\footnote{During the processing of this paper it was pointed 
out to the author that a fourth proof method can be found in the \cite{Franklin}.}. We will 
then show how the computation can be extended to the more general cases of non closed wires and non-uniform magnetic field.

\section{An elementary proof}\label{elementary_se}

The starting point is the Lorentz force acting on an element $\RMD\mbf{\ell}$ of the wire, which in SI units is
\begin{equation}\label{s0}
\RMD\mbf{F}=I \, \RMD\mbf{\ell}\times\mbf{B}\ ,
\end{equation}
where $I$ is the current. The torque acting on the wire is then
\begin{equation}\label{s1}
\mbf{M}=I\oint \mbf{r}\times(\RMD\mbf{\ell}\times\mbf{B})\ . 
\end{equation}
Let us now suppose the wire to be parametrized by the function $\mbf{s}(t)$, with $t\in [0,1]$ a real parameter. In this 
case $\RMD\mbf{\ell}=\dot{\mbf{s}}\RMD t$ (we denote by the dot the derivation with respect to $t$) and \Eqref{s1} can be 
rewritten in the form
\begin{equation}\label{s2}
\mbf{M}=I\int_0^1 \mbf{s}\times(\dot{\mbf{s}}\times\mbf{B})\RMD t\ . 
\end{equation}

By using the vectorial identity
\begin{equation}\label{sec1id}
\mbf{X}\times(\mbf{Y}\times\mbf{Z})=\mbf{Y}(\mbf{X}\cdot\mbf{Z})-\mbf{Z}(\mbf{X}\cdot\mbf{Y})
\end{equation}
we can rewrite \Eqref{s2} in the form
\begin{equation}\label{s3}
\mbf{M}=I\int_0^1 \Big\{\dot{\mbf{s}}(\mbf{s}\cdot\mbf{B})-\mbf{B}(\dot{\mbf{s}}\cdot\mbf{s})\Big\}\, \RMD t
\end{equation}
and it is simple to show that the second term vanishes in an uniform field: the vector 
$\mbf{B}$ can be carried out of the integral, which finally reduces to
\begin{equation}\label{s4}
\int_0^1 \dot{\mbf{s}}\cdot\mbf{s}\,\RMD t=
\frac{1}{2}\int_0^1 \frac{\RMD}{\RMD t}|\mbf{s}|^2 \, \RMD t =0
\end{equation}
since $\mbf{s}(0)=\mbf{s}(1)$ for a closed wire.

By expressing the first term of \Eqref{s3} in components we have (summation on the repeated indices is always 
assumed when not otherwise stated)
\begin{equation}\label{s5}
\eqalign{
& \int_0^1\Big[\dot{\mbf{s}}(\mbf{s}\cdot\mbf{B})\Big]_i\, \RMD t=B_j\int_0^1 \dot{s}_i s_j \, \RMD t= \\
& =\frac{1}{2}B_j\int_0^1(\dot{s}_i s_j+\dot{s}_j s_i)\RMD t+\frac{1}{2}B_j\int_0^1(\dot{s}_i s_j - \dot{s}_j s_i)\RMD t=\\
& =\frac{1}{2}B_j\int_0^1\frac{\RMD}{\RMD t}(s_is_j)\RMD t+\frac{1}{2}B_j\int_0^1(\dot{s}_i s_j - \dot{s}_j s_i)\RMD t=\\
& =\frac{1}{2}\int_0^1 \Big[ \dot{\mbf{s}}(\mbf{B}\cdot\mbf{s})-\mbf{s}(\mbf{B}\cdot\dot{\mbf{s}})\Big]_i \RMD t 
}
\end{equation}
where the total derivative term vanishes for the same reason as the integral in \Eqref{s4} does.  On the other hand 
from the identity in \Eqref{sec1id} we have
\begin{equation}
(\mbf{s}\times\dot{\mbf{s}})\times\mbf{B}=\dot{\mbf{s}}(\mbf{B}\cdot\mbf{s})-\mbf{s}(\mbf{B}\cdot\dot{\mbf{s}})\ ,
\end{equation}
so that from \Eqref{s3}, \eqref{s4} and \eqref{s5} we get
\begin{equation}
\mbf{M}=\left(\frac{I}{2}\int_0^1 (\mbf{s}\times\dot{\mbf{s}})\, \RMD t\right) \times\mbf{B}\ ,
\end{equation}
which is the desired \Eqref{torque} with the identification of the dipole magnetic moment
\begin{equation}\label{mu1}
\mbf{\mu}=\frac{I}{2}\int_0^1 \mbf{s}\times\dot{\mbf{s}}\, \RMD t\ .
\end{equation}
Going back to the line integral form we finally obtain
\begin{equation}\label{mu2}
\mbf{\mu}=\frac{I}{2}\oint \mbf{r}\times\RMD\mbf{\ell}\ ,
\end{equation}
which for planar wires reduces to the simple expression $AI\mbf{n}$, since the element of area is given by 
$\mbf{n}\RMD A=\frac{1}{2}\mbf{r}\times\RMD\mbf{\ell}$.

\section{Landau's proof}\label{landau_sec}

In this section we will present a proof of \Eqref{torque} by using the identity \Eqref{landau_id}, whose proof is 
given in \Appref{sec_stokes}.

We first of all show how the form \Eqref{mu2} of the magnetic dipole moment can be simplified by using the extension 
of the Stokes theorem proven in the appendix: by using \Eqref{landau_id} we immediately get (since
$\mbf{\nabla}\times\mbf{r}=0$, $(\mbf{a}\cdot\mbf{\nabla})\mbf{r}=\mbf{a}$ and $\mbf{\nabla}\cdot\mbf{r}=3$)
\begin{equation}
\eqalign{
\oint\RMD\mbf{\ell}\times\mbf{r}&=\int\RMD\mbf{\sigma}\times (\mbf{\nabla}\times \mbf{r})+
\int(\RMD\mbf{\sigma}\cdot\mbf{\nabla})\mbf{r}-\int\RMD\mbf{\sigma} (\mbf{\nabla}\cdot\mbf{r})=\\
&= \int\RMD\mbf{\sigma}-3\int\RMD\mbf{\sigma}=-2\int\RMD\mbf{\sigma}
}
\end{equation}
and thus \Eqref{mu2} becomes
\begin{equation}\label{mu3}
\mbf{\mu}=I\int\RMD\mbf{\sigma}\ ,
\end{equation}
which is the simplest extension to non planar wires of the expression $\mbf{\mu}=AI\mbf{n}$ valid in the planar 
case. Clearly the result of \Eqref{mu3} does not depend on the choice of the surface of integration: if we denote 
by $\mbf{c}$ a constant vector, the difference between the (projection on $\mbf{c}$ of the) results obtained with 
two different choices $\Sigma_1$ and $\Sigma_2$ is given by
\begin{equation}
\mbf{c}\cdot(\mbf{\mu}_1-\mbf{\mu}_2)=I\left(\int_{\Sigma_1}\RMD\mbf{\sigma}\cdot\mbf{c}-\int_{\Sigma_2}\RMD\mbf{\sigma}
\cdot\mbf{c}\right)
=I\int_{V_{12}} (\mbf{\nabla}\cdot\mbf{c})\,\RMD^3\mbf{r}=0\ ,
\end{equation}
where $V_{12}$ is the volume bounded by the surfaces $\Sigma_1$ and $\Sigma_2$. Since this is true for every $\mbf{c}$ we 
conclude that $\mbf{\mu}_1=\mbf{\mu}_2$.

In order to apply \Eqref{landau_id} to the computation of the torque it is convenient to use the following vectorial 
identity
\begin{equation}\label{sec2id}
\mbf{X}\times(\mbf{Y}\times\mbf{Z})+\mbf{Y}\times (\mbf{Z}\times\mbf{X})+\mbf{Z}\times(\mbf{X}\times\mbf{Y})=0
\end{equation}
and rewrite \Eqref{s1} in the form
\begin{equation}\label{s6}
\mbf{M}=-I\oint\RMD\mbf{\ell}\times(\mbf{B}\times\mbf{r})+I\oint (\mbf{r}\times\RMD\mbf{\ell})\times\mbf{B}\ .
\end{equation}
By comparison with \Eqref{mu2} the second term can be recast in the form $2\mbf{\mu}\times\mbf{B}$, while applying 
\Eqref{landau_id} to the first term we get
\begin{equation}
\eqalign{
& I\oint \RMD\mbf{\ell}\times(\mbf{B}\times\mbf{r})=
I\int\RMD\mbf{\sigma}\times\Big(\mbf{\nabla}\times(\mbf{B}\times\mbf{r})\Big)+\\
& +I\int (\RMD\mbf{\sigma}\cdot \mbf{\nabla})(\mbf{B}\times\mbf{r})-I\int\RMD\mbf{\sigma}\Big(\mbf{\nabla}
\cdot(\mbf{B}\times\mbf{r})\Big)\ .
}
\end{equation} 
By using the relations (which can be easily checked by direct computation) 
\begin{equation}
\eqalign{
& \mbf{\nabla}\times(\mbf{B}\times\mbf{r})=2\mbf{B}\\
& (\RMD\mbf{\sigma}\cdot\mbf{\nabla})(\mbf{B}\times\mbf{r})=\mbf{B}\times\RMD\mbf{\sigma}\\
& \mbf{\nabla}\cdot(\mbf{B}\times\mbf{r})=0
}
\end{equation}
and \Eqref{mu3} we thus get
\begin{equation}
I\oint \RMD\mbf{\ell}\times(\mbf{B}\times\mbf{r})=2I\int\RMD\mbf{\sigma}\times\mbf{B}
+I\int\mbf{B}\times\RMD\mbf{\sigma}= I\int \RMD\mbf{\sigma}\times \mbf{B}=\mbf{\mu}\times\mbf{B}\ .
\end{equation}
Using this result in \Eqref{s6} we finally obtain \Eqref{torque}.

\section{Jackson's proof}\label{jackson_sec}

This proof makes use of \Eqref{jackson_id}, which is easily proven: since we assumed $\mbf{Y}$ to be a function of compact
support we have, by using the divergence theorem for a large enough volume $V$ (bounded by the surface $\Sigma$), 
\begin{equation}
\int_V \mbf{\nabla}\cdot(fg\mbf{Y})\RMD^3\mbf{r}=\int_{\Sigma} fg\,\RMD\mbf{\sigma}\cdot{\mbf{Y}}=0\ ,
\end{equation}
since $\mbf{Y}$ vanishes on $\Sigma$. By using the identity
\begin{equation}
\mbf{\nabla}\cdot(fg\mbf{Y})=f\mbf{Y}\cdot\mbf{\nabla}g+g\mbf{Y}\cdot\mbf{\nabla}f+fg\mbf{\nabla}\cdot\mbf{Y}
\end{equation}
we thus get \Eqref{jackson_id}.

The starting point is \Eqref{s1}, which can be rewritten by noting that the current density $\mbf{j}$ has support on the 
wire and that $I\RMD\mbf{\ell}=\mbf{j}\RMD^3\mbf{r}$, thus
\begin{equation}
\mbf{M}=I\oint \mbf{r}\times(\RMD\mbf{\ell}\times\mbf{B})=\int \mbf{r}\times(\mbf{j}\times\mbf{B})\RMD^3\mbf{r}\ .
\end{equation}
By using \Eqref{sec1id} this expression becomes
\begin{equation}\label{s7}
\mbf{M}=\int \mbf{j}(\mbf{r}\cdot\mbf{B})\RMD^3\mbf{r}-\int \mbf{B}(\mbf{r}\cdot\mbf{j})\RMD^3\mbf{r}\ .
\end{equation}

If we now use \Eqref{jackson_id} with $f=g=r_i$ (component $i$ of $\mbf{r}$) and $\mbf{Y}=\mbf{j}$, remembering that 
$\mbf{\nabla}\cdot\mbf{j}=0$, we get (no summation over repeated indices)
\begin{equation}
0=\int (f\mbf{Y}\cdot\mbf{\nabla}g+g\mbf{Y}\cdot\mbf{\nabla}f)\RMD^3\mbf{r}=2\int r_ij_i\, \RMD^3\mbf{r}
\end{equation}
and by summing over $i$ we obtain
\begin{equation}
\int \mbf{r}\cdot\mbf{j}\,\RMD^3\mbf{r}=0\ ,
\end{equation}
so the second term of \Eqref{s7} vanishes in an uniform magnetic field.

By using instead $f=r_i$, $g=r_k$ and $\mbf{Y}=\mbf{j}$ we get (again no summation on indices)
\begin{equation}
\int (r_ij_k+r_kj_i)\RMD^3\mbf{r}=0
\end{equation}
and, by means of manipulations analogous to the ones in \Eqref{s5}, we obtain
\begin{equation}
\int \mbf{j}(\mbf{r}\cdot\mbf{B})\RMD^3\mbf{r}=\frac{1}{2}\int\Big(\mbf{j}(\mbf{B}\cdot\mbf{r})-\mbf{r}(\mbf{B}\cdot
\mbf{j})\Big)\RMD^3\mbf{r}
=\frac{1}{2}\int (\mbf{r}\times\mbf{j})\times\mbf{B}\RMD^3\mbf{r}\ ,
\end{equation}
so that
\begin{equation}\label{s8}
\mbf{M}=\frac{1}{2}\int (\mbf{r}\times\mbf{j})\times\mbf{B}\RMD^3\mbf{r}\ .
\end{equation}

By using again the fact that $I\RMD\mbf{\ell}=\mbf{j}\RMD^3\mbf{x}$ we see by comparison with \Eqref{mu2} that
\begin{equation}\label{mu4}
\mbf{\mu}=\frac{1}{2}\int\mbf{r}\times\mbf{j}\RMD^3\mbf{r}
\end{equation}
and \Eqref{s8} thus reduces to \Eqref{torque}.

\section{Comments and extensions}

We have seen in the previous sections that the torque generated by an uniform magnetic field $\mbf{B}$ on a wire is given by
\begin{equation}\label{torque2}
\mbf{M}=\mbf{\mu}\times\mbf{B}\ ,
\end{equation}
where the magnetic dipole moment $\mbf{\mu}$ has the following equivalent definitions:
\begin{equation}\label{s9}
\mbf{\mu}=\frac{I}{2}\int_0^1 \mbf{s}\times\dot{\mbf{s}}\, \RMD t=
\frac{I}{2}\oint \mbf{r}\times\RMD\mbf{\ell}
=I\int\RMD\mbf{\sigma}=\frac{1}{2}\int\mbf{r}\times\mbf{j}\RMD^3\mbf{r}\ .
\end{equation}

It is instructive to go through the different proofs presented of the relation \Eqref{torque2} to search for the key 
ingredients used: from the first proof it is clear that, for a formula like \Eqref{torque2} to be valid, the wire must be 
closed, an aspect whose importance is not completely evident in the usual example of the rectangular wire. This same 
requirement is naturally fundamental also for the second proof, since the Stokes theorem could not be applied to an open 
wire, and, although in a less trivial way, it is fundamental also in the third proof: there the main ingredient was current 
conservation, but the current would not be conserved in an open wire (see also below). Also, in all the proofs, the 
uniformity of the magnetic field was crucial to carry $\mbf{B}$ out of the integration and factorize $\mbf{\mu}$.

\begin{figure}[t]
\centering
\scalebox{0.35}{\rotatebox{-90}{\includegraphics*{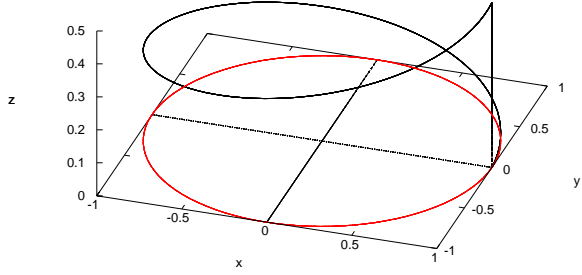}}}
\caption{The wire parametrically represented by \Eqref{s10} with $\alpha=1$.}\label{spiral_fig}
\end{figure}

The first form of \Eqref{s9} is usually the most direct one to use in computations involving non-planar wires which are 
not trivially decomposable into planar ones. A simple non-trivial example is the wire in \Figref{spiral_fig}, whose 
parametrization is
\begin{equation}\label{s10}
\mbf{s}(t)=\left\{\begin{array}{ll}
\cos(4\pi t)\hat{x}+\sin(4\pi t)\hat{y}+\alpha t\hat{z} & t\in[0,1/2] \\
\rule{0.0cm}{0.5cm}\hat{x}+\alpha (1-t)\hat{z} & t\in(1/2,1] \end{array} \right. \ .
\end{equation}
It is simple to show that
\begin{equation}
\int_0^1\mbf{s}\times\dot{\mbf{s}}\,\RMD t=\alpha\hat{y}+2\pi\hat{z}
\end{equation}
and thus the magnetic dipole moment of the wire parametrized by \Eqref{s10} is
\begin{equation}
\mbf{\mu}=\frac{I\alpha}{2}\hat{y}+I\pi\hat{z}\ ,
\end{equation}
which clearly reduce to the planar result $AI\hat{z}$, with $A=\pi$, when $\alpha=0$.

Various extensions of the result in \Eqref{torque2}-\eqref{s9} can easilly be performed: a particularly simple one is to 
consider a non-closed wire, i.e. $\mbf{\Delta}=\mbf{s}(1)-\mbf{s}(0)\neq 0$. Clearly in this case on the wire it is acting 
also a net force: from \Eqref{s0} it is immediate to get for the total force the expression
\begin{equation}
\mbf{F}=I\mbf{\Delta}\times \mbf{B}
\end{equation}
and collecting the previously vanishing terms in \Eqref{s4}-\eqref{s5} we obtain for the torque
\begin{equation}\label{s11}
\eqalign{
\mbf{M}&=\mbf{\mu}\times\mbf{B}+\frac{I}{2}\big(\mbf{s}(1)(\mbf{s}(1)\cdot\mbf{B})-\mbf{B}|\mbf{s}(1)|^2\big)-\\
&\qquad -\frac{I}{2}\big(\mbf{s}(0)(\mbf{s}(0)\cdot\mbf{B})-\mbf{B}|\mbf{s}(0)|^2\big)=\\
&=\mbf{\mu}\times\mbf{B}+\frac{I}{2}\mbf{s}(1)\times(\mbf{s}(1)\times\mbf{B})-\\
&\qquad -\frac{I}{2}\mbf{s}(0)\times(\mbf{s}(0)\times\mbf{B})\ ,
}
\end{equation}
where now $\mbf{\mu}$ is defined by the first expression in \Eqref{s9}. Since a net force is acting on the wire, the 
torque depends on the choice of the pole used (i.e. on the origin of the coordinates in our computation) and \Eqref{s11} 
can not in general be written in term of $\mbf{\Delta}$ only; this happens only if we choose the pole in $\mbf{s}(0)$, in 
which case \Eqref{s11} collapses to
\begin{equation}
\mbf{M}=\mbf{\mu}\times\mbf{B}+\frac{I}{2}\mbf{\Delta}\times(\mbf{\Delta}\times\mbf{B})
\quad (\,\mathrm{pole\ in\ }\mbf{s}(0)\,)\ .
\end{equation} 
This result can be obtained also using the methods of \Secref{jackson_sec} by noting that, for a non-closed wire, the 
current is not conserved and a source and a sink have to be present at the wire endings:
\begin{equation}
\mbf{\nabla}\cdot\mbf{j}=I\delta(\mbf{r}-\mbf{s}(0))-I\delta(\mbf{r}-\mbf{s}(1))\ .
\end{equation}

Another possible extension is the one to a non uniform magnetic field (again for a closed wire). Let us consider for 
simplicity only the first linear correction to the uniform field case:
\begin{equation}\label{s12}
\mbf{B}(\mbf{r})=\mbf{b}+\underline{\mbf{a}}\mbf{r}\ ,
\end{equation}
where $\mbf{b}$ is a constant vector and $\underline{\mbf{a}}$ is a linear operator, i.e. in component we have
\begin{equation}
B_i(\mbf{r})=b_i+a_{ij}r_j\ .
\end{equation}
The requirement $\mbf{\nabla}\cdot\mbf{B}=0$ imposes the restriction $\mathrm{Tr}\underline{\mbf{a}}=0$ and, if we 
further assume that the currents that generate $\mbf{B}$ are far away from the wire (``far away'' means here that 
these currents do not contribute to the various line or surface integrals), from $\mbf{\nabla}\times\mbf{B}=0$ the 
relation $a_{ij}=a_{ji}$ follows, i.e. the matrix $\underline{\mbf{a}}$ is symmetric.
Since \Eqref{s3} is linear in $\mbf{B}$, we can calculate the corrective term to $\mbf{M}$ by simply using $\mbf{b}=0$. 
We than have (using $\mbf{s}(0)=\mbf{s}(1)$)
\begin{equation}
\eqalign{
&\int_0^1\Big[\mbf{B}(\dot{\mbf{s}}\cdot\mbf{s})\Big]_i\RMD t=a_{ij}\int_0^1s_j\dot{s}_ks_k\RMD t=\\
&=\frac{1}{2}a_{ij}\int_0^1s_j\frac{\RMD}{\RMD t}(s_k)^2\RMD t =-\frac{1}{2}\int_0^1(s_k)^2a_{ij}\dot{s}_j\RMD t
}
\end{equation}
and thus
\begin{equation}
\int_0^1 \mbf{B}(\dot{\mbf{s}}\cdot\mbf{s})\RMD t=-\frac{1}{2}\oint r^2 \underline{\mbf{a}}\RMD\mbf{\ell}\ .
\end{equation}
On the other hand
\begin{equation}
\int_0^1 \dot{\mbf{s}}(\mbf{s}\cdot\mbf{B})\RMD t=\oint\RMD \mbf{\ell}\big(\mbf{r}\cdot(\underline{\mbf{a}}\mbf{r})\big) 
\end{equation}
and we thus obtain for the torque caused by the non-uniform magnetic field in \Eqref{s12} the expression
\begin{equation}\label{s13}
\mbf{M}=\mbf{\mu}\times\mbf{b}+\frac{I}{2}\oint r^2 \underline{\mbf{a}}\RMD\mbf{\ell}
-I\oint\RMD \mbf{\ell}\big(\mbf{r}\cdot(\underline{\mbf{a}}\mbf{r})\big)\ . 
\end{equation}
The second term can be written also as a surface integral, indeed if we use the result \Eqref{stokes2} of 
\Appref{sec_stokes2} with $\mbf{B}$ given by \Eqref{s12} we get
\begin{equation}\label{s13bis}
\mbf{M}=\mbf{\mu}\times\mbf{b}+I\int\RMD\mbf{\sigma}\times (\underline{\mbf{a}}\mbf{r})+I
\int\mbf{r}\times(\underline{\mbf{a}}\RMD\mbf{\sigma})
\end{equation}

Clearly in the non-uniform field case also a non-vanishing net force is in general present, which, by using 
\Eqref{landau_id} (remembering that $\mbf{\nabla}\cdot\mbf{B}=0$ and $\mbf{\nabla}\times\mbf{B}=0$), can be written as
\begin{equation}\label{s15}
\mbf{F}=\int(\RMD\mbf{\sigma}\cdot\mbf{\nabla})\mbf{B}=\underline{\mbf{a}}\mbf{\mu}=\mbf{\nabla}(\mbf{\mu}\cdot\mbf{B})
\end{equation}

If we denote by $b$ the modulus of $\mbf{b}$, by $a$ a typical value of $\underline{\mbf{a}}$ and we consider a wire of 
typical linear dimension $L$, the contributions in \Eqref{s13}-\eqref{s15} are of order
\begin{equation}\label{s16}
M\sim bL^2+aL^3 \qquad F\sim aL^2
\end{equation}
and the first of these equations can conveniently be rewritten as
\begin{equation}\label{s16bis}
M\sim b L^2\left(1+\frac{L}{\lambda}\right)\ ,
\end{equation} 
where $\lambda=b/a$ is the typical length scale of variation of the magnetic field in \Eqref{s12}. It is then clear that, as
intuitively obvious, the non uniformity of the magnetic field can be neglected as far as $L\ll \lambda$. For dimensional
reasons the force in \Eqref{s16} has one power of $L$ missing with respect to the torque, but dimensionality would suggests 
also the presence of a term $bL$, which is absent since in an uniform field no net force is acting on the wire. Because of 
the absence of this leading contribution, the non uniformity of the magnetic field can not be neglected even for small wires 
in the force computation.

For a generic non-uniform magnetic field, \Eqref{s12} is just the first term of a Taylor expansion, whose general form is
\begin{equation}\label{s17}
B_i=a^{(0)}_i+a^{(1)}_{ij_1}r_{j_1}+a^{(2)}_{ij_1j_2}r_{j_1}r_{j_2}+\cdots
+a^{(n)}_{ij_1\cdots j_n}r_{j_1}\cdots r_{j_n}+\cdots
\end{equation}
where $a^{(n)}_{ij_1\cdots j_n}$ is symmetric under permutations of $j_1,\ldots, j_n$. The condition 
$\mbf{\nabla}\cdot\mbf{B}=0$ becomes for the $n-$th term
\begin{equation}
0=\partial_i B_i=a^{(n)}_{ij_i\cdots j_n}\partial_i(r_{j_1}\cdots r_{j_n})
=na^{(n)}_{iij_2\cdots j_n}r_{j_2}\cdots r_{j_n}
\end{equation}
and the condition $\mbf{\nabla}\times\mbf{B}=0$ gives
\begin{equation}
0=\partial_{\alpha}B_{\beta}-\partial_{\beta}B_{\alpha}
=n(a^{(n)}_{\beta\alpha j_2\cdots j_n}-a^{(n)}_{\alpha\beta j_2\cdots j_n})r_{j_2}\cdots r_{j_n}\ ,
\end{equation}
so that $a^{(n)}_{ij_1\cdots j_n}$ is again completely symmetric and traceless. It is then not difficult to generalize 
\Eqref{s13} and \Eqref{s13bis}.
We can introduce a characteristic length $\lambda_{(n)}$ for every term in \Eqref{s17} by 
\begin{equation}
\lambda_{(n)}=\sqrt[n]{\frac{a^{(0)}}{a^{(n)}}}\ ,
\end{equation}
where $a^{(0)}$ and $a^{(n)}$ stand here for typical values, and \Eqref{s16bis} generalizes to
\begin{equation}\label{s18}
M\sim a^{(0)} L^2\Big[1+\frac{L}{\lambda_{(1)}}+\left(\frac{L}{\lambda_{(2)}}\right)^2+\cdots
+\left(\frac{L}{\lambda_{(n)}}\right)^n+\cdots\Big]\ .
\end{equation}
For a typical magnetic field we have
\begin{equation}
\lambda_{(1)}\lesssim \lambda_{(2)}\lesssim\cdots\lesssim\lambda_{(n)}\lesssim\cdots
\end{equation}
and we thus see that the expansion \Eqref{s18} can be truncated to the $n-$th term if the typical linear dimension of the 
wire satisfies
\begin{equation}
\frac{L}{\lambda_{(n+1)}}\left(\frac{\lambda_{(n)}}{\lambda_{(n+1)}}\right)^n\ll 1\ .
\end{equation}

\section{Conclusions}

We discussed three different methods to compute the torque acting on a generic wire in an uniform magnetic field, the first 
is completely elementary, the other two present a higher degree of mathematical sophistication.
We have then shown how the computation can be generalized to the cases of non-closed wires and non-uniform magnetic field.

\ack
It is a pleasure to acknowledge Paolo Christillin and Maurizio Fagotti for useful comments and discussions.

\appendix 

\section{An extension of the Stokes theorem}\label{sec_stokes}

The Stokes theorem relates the circuitation of a field along a closed curve to the flux of its curl: in formulae (see e.g. 
Ref.~\cite{rudin})
\begin{equation}\label{stokes}
\oint \RMD\mbf{\ell}\cdot \mbf{X}=\int\RMD\mbf{\sigma}\cdot (\mbf{\nabla}\times \mbf{X})\ ,
\end{equation}
while we are interested in line integrals of the form
\begin{equation}\label{I}
\mbf{I}=\oint\RMD\mbf{\ell}\times\mbf{X}\ .
\end{equation}

In order to make use of the Stokes theorem in the computation of the r.h.s of \Eqref{I}, it is convenient to take the scalar 
product of $\mbf{I}$ with a constant vector, which we will denote by $\mbf{c}$:
\begin{equation}\label{stokes_step1}
\mbf{c}\cdot\mbf{I}=\oint \mbf{c}\cdot(\RMD\mbf{\ell}\times\mbf{X})=\oint \RMD\mbf{\ell}\cdot (\mbf{X}\times\mbf{c})
=\int \RMD\mbf{\sigma}\cdot\Big[\mbf{\nabla}\times(\mbf{X}\times\mbf{c})\Big]\ ,
\end{equation}
where in the intermediate step we used the identity $\mbf{A}\cdot(\mbf{B}\times\mbf{C})=\mbf{B}\cdot(\mbf{C}\times\mbf{A})$ 
and the last identity is just the usual Stokes theorem.

Passing in components and remembering that $\mbf{c}$ is a constant vector, we get
\begin{equation}\label{epsilon}
\eqalign{
\Big[\mbf{\nabla}\times(\mbf{X}\times \mbf{c})\Big]_i&= \epsilon_{ijk}\partial_j\epsilon_{klm}X_lc_m=\\
&=-\epsilon_{kji}\epsilon_{klm}\partial_j X_l c_m=\\
&= -(\delta_{jl}\delta_{im}-\delta_{jm}\delta_{il})\partial_j X_l c_m
}
\end{equation}
and thus
\begin{equation}
\mbf{\nabla}\times(\mbf{X}\times \mbf{c})=-\mbf{c}(\mbf{\nabla}\cdot\mbf{X})+(\mbf{c}\cdot\mbf{\nabla})\mbf{X}\ .
\end{equation}
In a similar way it can be shown that 
\begin{equation}
\mbf{c}\times(\mbf{\nabla}\times\mbf{X})=-(\mbf{c}\cdot\mbf{\nabla})\mbf{X}+\mbf{\nabla}(\mbf{c}\cdot\mbf{X}) 
\end{equation}
and by summing theese two equations we get
\begin{equation}\label{stokes_step2}
\mbf{\nabla}\times (\mbf{X}\times\mbf{c})=(\mbf{\nabla}\times \mbf{X})\times\mbf{c}+\mbf{\nabla}(\mbf{c}\cdot\mbf{X})
-\mbf{c}(\mbf{\nabla}\cdot\mbf{X})\ .
\end{equation}
By using this identity in \Eqref{stokes_step1} we obtain 
\begin{equation}
\eqalign{
\mbf{c}\cdot\mbf{I}&=\int\RMD\mbf{\sigma}\cdot\Big\{(\mbf{\nabla}\times\mbf{X})\times\mbf{c}\Big\}+
\int\RMD\mbf{\sigma}\cdot\mbf{\nabla}(\mbf{c}\cdot\mbf{X})-\\
&\qquad -\int \mbf{c}\cdot\RMD\mbf{\sigma} (\mbf{\nabla}\cdot\mbf{X})=\\
&=\int\mbf{c}\cdot\Big\{\RMD\mbf{\sigma}\times (\mbf{\nabla}\times\mbf{X})\Big\}+
\int\RMD\mbf{\sigma}\cdot\mbf{\nabla}(\mbf{c}\cdot\mbf{X})-\\
&\qquad -\int \mbf{c}\cdot\RMD\mbf{\sigma} (\mbf{\nabla}\cdot\mbf{X})
}
\end{equation}
and by replacing the constant vector $\mbf{c}$ by the versors of the coordinate axes we finally get the desired extension of 
the Stokes theorem
\begin{equation}
\mbf{I}=\int\RMD\mbf{\sigma}\times (\mbf{\nabla}\times\mbf{X})+
\int(\RMD\mbf{\sigma}\cdot\mbf{\nabla})\mbf{X} -\int \RMD\mbf{\sigma} (\mbf{\nabla}\cdot\mbf{X})\ .
\end{equation}

\section{Another variant of the Stokes theorem}\label{sec_stokes2}

In this appendix we will deduce another variant of the Stokes theorem, this time referred to integrals of the form of 
the torque:
\begin{equation}
\mbf{J}=\oint \mbf{r}\times(\RMD\mbf{\ell}\times\mbf{B})
\end{equation}
Multiplying $\mbf{J}$ by the constant vector $\mbf{c}$ we can use
\begin{equation}
\mbf{c}\cdot\Big(\mbf{r}\times(\RMD\mbf{\ell}\times\mbf{B})\Big)
=(\RMD\mbf{\ell}\times\mbf{B})\cdot(\mbf{c}\times\mbf{r})=\RMD\mbf{\ell}\cdot\Big(\mbf{B}\times (\mbf{c}\times\mbf{r})\Big)
\end{equation}
to get, by the Stokes theorem,
\begin{equation}
\mbf{c}\cdot\mbf{J}=\int \RMD\mbf{\sigma}\cdot\Big\{\mbf{\nabla}\times\Big(\mbf{B}\times (\mbf{c}\times\mbf{r})\Big)\Big\}\ .
\end{equation}
By proceeding as in \Eqref{epsilon} it can be shown that
\begin{equation}
\mbf{\nabla}\times(\mbf{Y}\times\mbf{Z})=(\mbf{Z}\cdot\mbf{\nabla})\mbf{Y}-(\mbf{Y}\cdot\mbf{\nabla})\mbf{Z}
+\mbf{Y}(\mbf{\nabla}\cdot \mbf{Z})-\mbf{Z}(\mbf{\nabla}\cdot\mbf{Y})
\end{equation}
and, by using $\mbf{\nabla}\cdot \mbf{B}=0$ and $\mbf{\nabla}\cdot(\mbf{c}\times\mbf{r})=0$, we get
\begin{equation}\label{stokes2}
\mbf{c}\cdot\mbf{J}=\int\RMD\mbf{\sigma}\cdot\Big[\Big((\mbf{c}\times\mbf{r})\cdot\mbf{\nabla}\Big)\mbf{B}\Big]
+\mbf{c}\cdot\int\RMD\mbf{\sigma}\times\mbf{B}\ .
\end{equation}
For an uniform magnetic field only the last term survives and gives once again \Eqref{torque}.

\end{document}